\documentclass{article}
\usepackage{spconf,amsmath,graphicx,hyperref,amsfonts}

\usepackage{booktabs}
\usepackage{tabularx}
\title{MALEFA: Multi-grAnularity Learning and Effective False Alarm Suppression for Zero-Shot Keyword Spotting}
\name{Lo-Ya Li$^{*}$, Tien-Hong Lo$^{*}$,Jeih-Weih Hung$^{\dagger}$, Shih-Chieh Huang$^{\P}$, Berlin Chen$^{*}$ }
\address{
$^{*}$National Taiwan Normal University, Taiwan \\
$^{\dagger}$National Chi Nan University, Taiwan \\
$^{\P}$Realtek Semiconductor Corp, Taiwan \\
}
\ninept
\begin{document}
\maketitle
%
\begin{abstract}
User-defined keyword spotting (KWS) without resorting to domain-specific pre-labeled training data is of fundamental importance in building adaptable and personalized voice interfaces. However, such systems are still faced with arduous challenges, including constrained computational resources and limited annotated training data. Existing methods also struggle to distinguish acoustically similar keywords, often leading to a pesky false alarm rate (FAR) in real-world deployments. To mitigate these limitations, we put forward MALEFA, a novel lightweight zero-shot KWS framework that jointly learns utterance- and phoneme-level alignments via cross-attention and a multi-granularity contrastive learning objective. Evaluations on four public benchmark datasets show that MALEFA achieves a high accuracy of 90\%, significantly reducing FAR to 0.007\% on the AMI dataset. Beyond its strong performance, MALEFA demonstrates high computational efficiency and can readily support real-time deployment on resource-constrained devices.
\end{abstract}
%
\section{Introduction}
\label{sec:intro}
Keyword spotting (KWS) enables intuitive human-computer interaction, facilitating the activation of voice assistants or smart devices with spoken commands, especially in hands-busy situations such as driving or gaming. Conventional KWS systems typically operate under a closed-set paradigm (using predefined wake words like “Hey Siri”, “OK Google” or others) and rely on extensive pre-defined training data \cite{sainath2015cnn, chen2014small, deep}. While effective in controlled conditions, their fixed-vocabulary setting limits the adaptability to user-defined or previously unseen keywords, hindering personalization in realistic use cases.
To overcome these restrictions, zero-shot KWS (ZSKWS) has emerged as a promising alternative \cite{wei2021open, zero, yusuf21}, enabling the detection of arbitrary spoken keywords in given speech segments based merely on matching with their textual representations. This paradigm eliminates the need for keyword-specific audio data for training or fine-tuning, making it more geared to practical deployment. 
\begin{figure}[t]
    \centering
    \includegraphics[width=1\linewidth]{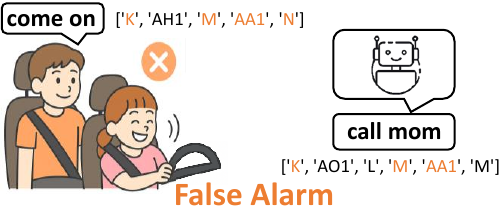}
    \caption{An example of false alarm triggered by phonetic similarity between keywords.}
    \label{fig:FA}
\end{figure}
One prominent ZSKWS approach is the cross-modal framework (CMCD) \cite{shin2022learning}, which aligns audio utterances and textual queries in a shared embedding space, allowing for flexible keyword matching and obviating the reliance on keyword-specific audio examples. Building on this, several studies have proposed methods to enhance alignment accuracy and generalization \cite{phonmatchnet, u2kws, adakws}. Despite these advances, existing systems rely on coarse-grained global representations of spoken utterances, narrowing down their ability to resolve phonetically similar keyword pairs. As illustrated in Figure~\ref{fig:FA}, semantically distinct keywords often share overlapping phonetic content (highlighted in orange), leading to high false alarms (FA). Meanwhile, MM-KWS \cite{mmkws} and CED \cite{ced} utilize Conformer \cite{conformer} to facilitate cross-lingual feature robustness. However, the substantial computational demands of these large pre-trained audio encoders hinder real-time deployment on resource-constrained devices. Therefore, there is a need for lightweight, robust alternatives that can effectively address phonetically ambiguous keywords while maintaining efficiency in low-resource scenarios.

In light of this, we propose MALEFA\footnote{Implementation code : https://github.com/Debbyyy10158/MALEFA}, a multi-granularity contrastive learning framework for ZSKWS. The proposed approach jointly learns utterance- and phoneme-level alignments, integrating cross-modal and contrastive learning techniques \cite{clap, clad, CLIP, plcl} for improved ZSKWS. MALEFA is specifically designed to reduce FA, maintaining high detection accuracy under computational constraints. 

Our main contributions are at least three-fold:
\begin{itemize}
\item \textbf{Multi-granularity contrastive learning:} A unified framework combining utterance- and phoneme-level contrastive objectives to capture both the global semantics of keywords and their fine-grained pronunciation.
\item \textbf{False alarm–aware loss:} We propose a better-tailored loss that directly penalizes false positives (FP) through a sigmoid-based precision constraint, explicitly optimizing for a low false alarm rate (FAR) on ZSKWS tasks.
\item \textbf{Lightweight on-device deployment:} Our model achieves FAR of 0.007\% and accuracy of 90\% on public benchmarks, with just 650K parameters and 93M FLOPs. 
\end{itemize}

\begin{figure*}[t]
    \centering
    \vspace{-20pt}  
    \includegraphics[width=0.7\textwidth, trim=0cm 0cm 0cm 0cm, clip]{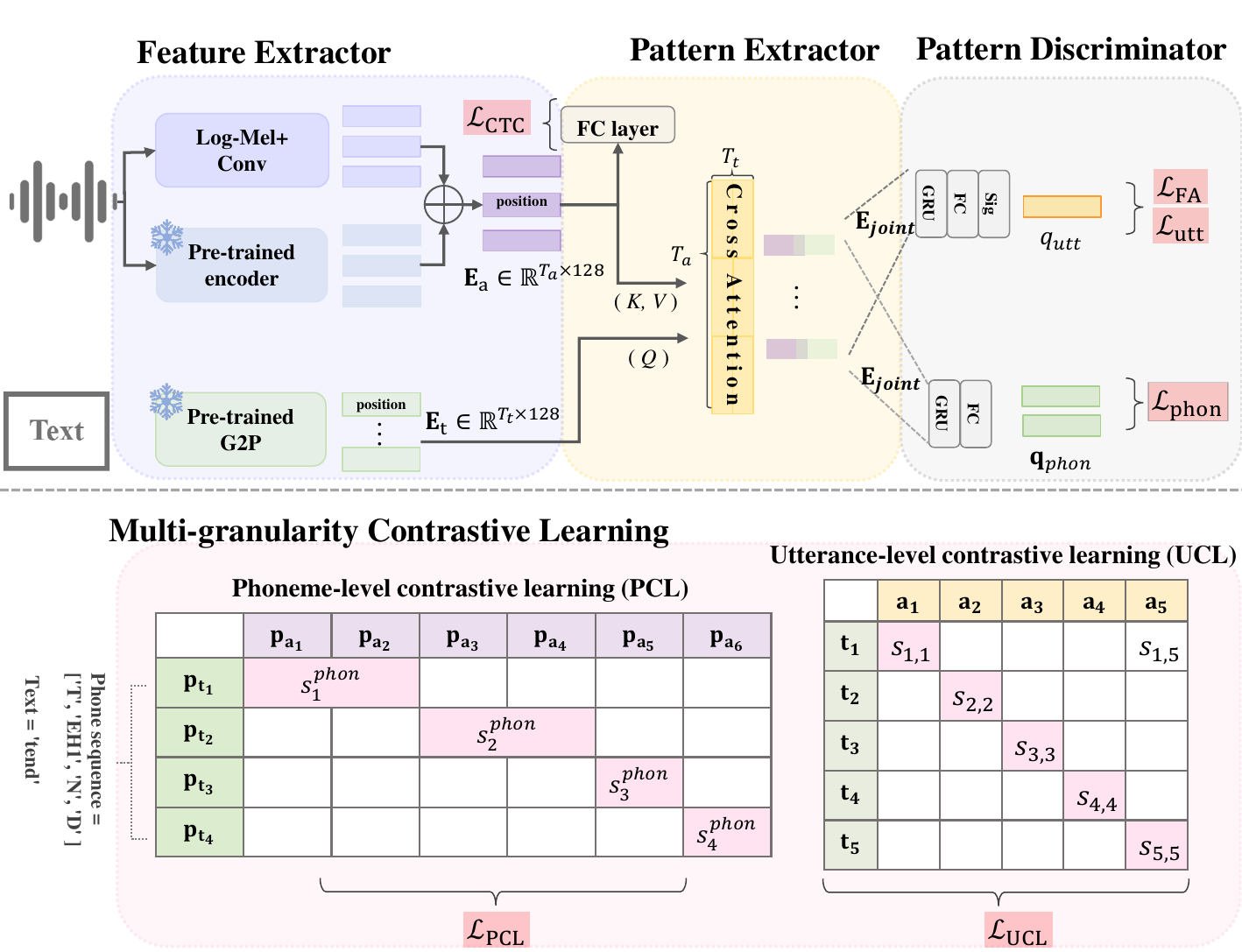}
    \vspace{-10pt}  
    \caption{Architecture of the proposed MALEFA model comprises three core components: (1) Feature Extractor (audio and text encoders), (2) Pattern Extractor (cross-attention module aligning text ($\text{Q}$) and audio ($\text{K,V}$) embeddings), and (3) Pattern Discriminator that produces both utterance-level and phoneme-level predictions. The model also incorporates detailed phoneme-level and utterance-level contrastive learning objectives, which are jointly optimized to resolve ambiguity in audio-to-text alignment.
    }
    \label{fig:architecture}
\end{figure*}

\begin{table*}[t]
\vspace{-15pt}
\centering
\caption{Comparison of MALEFA with prior ZSKWS models and ablation variants. 
Metrics include AUC (\%$\uparrow$), EER (\%$\downarrow$), and ACC$_4$ (\%$\uparrow$) on Google Speech Commands (G), Qualcomm (Q), LibriPhrase Easy (L$_E$), and LibriPhrase Hard (L$_H$). 
The full MALEFA achieves the best overall performance with only 0.7M parameters, while removing PCL, UCL, or FA-aware loss degrades accuracy (ACC) or increases equal error rates (EER), confirming their complementary contributions.}
\label{tab:performance}
\small
\newcolumntype{C}{>{\centering\arraybackslash}X}
\begin{tabularx}{\textwidth}{@{} l | CCCC | CCCC | C | C @{}}
\toprule
\textbf{Method} 
& \multicolumn{4}{c|}{AUC (\%){$\uparrow$}} 
& \multicolumn{4}{c|}{EER (\%){$\downarrow$}}
& ACC$_4$ (\%){$\uparrow$} 
& \# Params \\
\cmidrule(lr){2-5} \cmidrule(lr){6-9} \cmidrule(lr){10-10} \cmidrule(lr){11-11}
 & G & Q & LP$_E$ & LP$_H$ & G & Q & LP$_E$ & LP$_H$ & Q & \\
\midrule
CMCD \cite{shin2022learning} & 81.06 & 94.51 & 96.70 & 73.58 & 27.25 & 12.15 & 8.42 & 32.90 & -- & -- \\
PhonMatchNet \cite{phonmatchnet}* & 98.11 & 98.90 & 99.29 & 88.52 & 6.77  & 4.75  & 2.80 & 18.82 & 80.45 & 0.7M \\
CED \cite{ced}  & --    & --    & 99.84 & 92.70 & --    & --    & 1.70 & 14.70 & --    & 4.6M \\
CLAD \cite{clad}  & --    & --    & 97.03 & 76.15 & --    & --    & 8.65 & 30.30 & --    & 2.2M \\
ADML \cite{adml} & --    & --    & 99.86 & 88.71 & --    & --    & 1.33 & 20.09 & --    & 1.8M \\
\midrule
\textbf{Ours} & 99.13 & 99.81 & \textbf{99.98} & \textbf{93.58} & 3.88 & 1.92 & \textbf{1.14} & \textbf{13.91} & \textbf{98.77} & 0.7M \\
\midrule
\quad w/o PCL & \textbf{99.41} & \textbf{99.91} & 99.42 & 87.64 & \textbf{3.82} & \textbf{1.22} & 2.63 & 20.29 & 91.80 & 0.7M \\
\qquad w/o UCL  & 98.72 & 99.75 & 99.61 & 88.06 & 4.78  & 2.29  & 2.13 & 19.90 & 98.76 & 0.7M \\
\qquad\qquad w/o FA  & 94.83 & 97.57 & 96.07 & 86.47 & 9.85  & 8.16  & 8.62 & 21.10 & 84.19 & 0.7M \\
\bottomrule
\end{tabularx}
\vspace{-10pt}

\end{table*}

\begin{table}[t]
\centering
\caption{False alarm rate (FAR, \%$\downarrow$) comparison on AMI, Google Speech Commands (G), and Qualcomm (Q). Compared with PhonMatchNet, MALEFA reduces FAR by several orders of magnitude, achieving near-zero false alarms across all datasets. 
}
\label{tab:fasle alarm}
\small 
\newcolumntype{Y}{>{\centering\arraybackslash}X} 
\begin{tabularx}{\columnwidth}{@{} p{3.2cm} | YYY @{}} 
\toprule
\textbf{Method} 
& \multicolumn{3}{c}{FAR (\%){$\downarrow$}} \\
\cmidrule(lr){2-4}
 & AMI & G & Q  \\
\midrule
PhonMatchNet \cite{phonmatchnet}* & 17.879 & 7.438 & 5.743 \\
\midrule
\textbf{Ours} & \textbf{0.007} & \textbf{0.002} & \textbf{0.000} \\
\midrule
\quad w/o PCL & 0.085 & 0.019 & 0.105 \\
\qquad w/o UCL  & 1.334 & 3.580 & 0.029 \\
\qquad\qquad w/o FA  & 14.542 & 6.710 & 0.690 \\
\bottomrule
\end{tabularx}
\vspace{-10pt}

\end{table}

\section{Methodology}
\subsection{Feature Extractor}\label{sec:feature}
As schematically depicted in Fig.~\ref{fig:architecture}, MALEFA employs a two-stream encoder with separate audio and text encoders. 
Both audio and text modalities are processed independently and later aligned in the pattern extractor.

\noindent
\textbf{Audio encoder.}
Each utterance is passed through a pre-trained speech encoder~\cite{googleembedder} using a 775\,ms window with 80\,ms shift, producing 96-dimensional features.
In parallel, the raw waveform is converted into a log-mel spectrogram (25\,ms frame, 10\,ms hop), subsequently projected by a lightweight trainable convolution layer. 
The two feature streams are concatenated to form the audio embedding $\mathbf{E}_a \in \mathbb{R}^{T_a \times 128}$ of the utterance, where $T_a$ is the number of frames. The experimental setup is identical to \cite{phonmatchnet}.

\noindent
\textbf{Text encoder.} 
Keywords are first converted into phoneme sequences via a G2P converter~\cite{g2p}, and each phoneme is embedded by a fully connected layer with ReLU activation, yielding $\mathbf{E}_t \in \mathbb{R}^{T_t \times 128}$, where $T_t$ is the sequence length. 
Both $\mathbf{E}_a$ and $\mathbf{E}_t$ are augmented with sinusoidal positional encodings to capture temporal order and improve alignment robustness.
\subsection{Pattern Extractor}\label{sec:pattern}
As illustrated in Fig.~\ref{fig:architecture}, the pattern extractor employs cross-attention to align audio and text embeddings. 
The text embedding $\mathbf{E}_t$ serves as the query ($Q$), while the audio embedding $\mathbf{E}_a$ provides both keys and values ($K,V$). 
This allows each phoneme to attend to the most relevant audio frames, yielding a joint representation:
\begin{equation}
\mathbf{E}_{\text{joint}} = \text{CrossAttention}(Q=\mathbf{E}_t, K=\mathbf{E}_a, V=\mathbf{E}_a),
\end{equation}
where $\mathbf{E}_a \in \mathbb{R}^{T_a \times 128}$ and $\mathbf{E}_t \in \mathbb{R}^{T_t \times 128}$.

\subsection{Pattern Discriminator}\label{sec:discriminator}
The joint embedding $\mathbf{E}_{\text{joint}}$ is passed through a Gated Recurrent Unit (GRU), followed by two classification heads. 
One head predicts utterance-level matching probability score between audio and text, denoted as $q_{\text{utt}}$. While the other operates on temporal segments of $\mathbf{E}_{\text{joint}}$ to capture phoneme-level alignment sequence $\mathbf{q}_{\text{phon}}$.

\subsection{Multi-granularity Contrastive Learning}\label{sec:cl}
\textbf{Phoneme-level Contrastive Learning (PCL).} 
The audio encoder outputs frame-level CTC logits $\mathbf{z} \in \mathbb{R}^{T_a \times V}$, supervised by the standard CTC loss \cite{CTC}:
\begin{equation}
\mathcal{L}_{\text{CTC}} = -\log q_{\text{CTC}}(\mathbf{y}\mid\mathbf{z}),
\label{eq:ctc}
\end{equation}
where $q_{\text{CTC}}$ marginalizes over valid frame–phoneme alignments. 
With the aid of Viterbi decoding, we obtain alignment confidences $s_i$ for each audio-text pair. 
The corresponding PCL loss is 
\begin{equation}
\mathcal{L}_{\text{PCL}} = \tfrac{1}{N} \sum_{i=1}^{N} \big[m_i(1-s_i)^2 + (1-m_i)s_i^2\big],
\label{eq:pcl}
\end{equation}
where $m_i\in\{0,1\}$ indicates whether the $i$-th pair is matched. 
This encourages high alignment confidence for positives and penalizes spurious overlaps instead.

\noindent
\textbf{Utterance-level Contrastive Learning (UCL).} 
For a mini-batch of $M$ pairs, we compute a similarity matrix $S_{\text{utt}}\in\mathbb{R}^{M\times M}$. 
Text-to-audio ($s^{\text{text}}_{v,r}$) and audio-to-text ($s^{\text{audio}}_{v,r}$) scores are optimized bidirectionally:
\begin{equation}
\mathcal{L}_{\text{UCL}} = \tfrac{1}{2}(\ell_{\text{text}}+\ell_{\text{audio}}),
\label{eq:ucl}
\end{equation}
with each term defined by
\begin{equation}
\ell_{*} = -\tfrac{1}{M}\sum_{v=1}^{M}\sum_{r=1}^{M}\big[m_{v,r}\log\sigma(s^{*}_{v,r})
+(1-m_{v,r})\log(1-\sigma(s^{*}_{v,r}))\big].
\end{equation}
Here $m_{v,r}=1$ if audio $v$ matches text $r$, and $0$ otherwise. 
A mini-batch size of $M=5$ balances stability and discrimination.
\subsection{False Alarm-aware Loss}\label{sec:FA}
False alarms (FA), i.e., false keyword detections on non-target audio, remain a major challenge facing KWS. 
Conventional BCE training maximizes overall accuracy but does not explicitly penalize FA, often requiring post-hoc threshold tuning. We therefore introduce a precision-constrained objective:
\begin{equation}
\mathcal{L}_{\text{FA}} = -\log(\text{Precision}) 
+ \lambda \cdot \max(0, \alpha - \text{Precision}),
\label{eq:fa}
\end{equation}
where the first term discourages low-precision predictions and the second enforces a margin constraint if precision falls below $\alpha$ (scaled by $\lambda$) ~\cite{FA}. 
For differentiability, true positives (TP) and false positives (FP) are approximated by smooth sigmoid bounds:
\begin{align}
\text{TP} &= \sum (1 + \gamma\delta)\,\sigma(\gamma x - \delta)\,x_{\text{true}}, \\
\text{FP} &= \sum (1 + \gamma\delta)\,\sigma(\gamma x + \delta)\,(1 - x_{\text{true}}),
\end{align}
with $x_{\text{true}}\in\{0,1\}$, sigmoid $\sigma(\cdot)$, steepness $\gamma=7.0$, and offset $\delta=0.035$. 
We set $\alpha=0.9$ and $\lambda=10.0$. 
This auxiliary loss is combined with BCE to improve FA suppression during training.
\subsection{Training Criterion}
Both utterance- and phoneme-level predictions are supervised using BCE losses, denoted by 
$\mathcal{L}_{\text{utt}}$ and $\mathcal{L}_{\text{phon}}$, respectively. 
The overall training objective combines all loss terms:
\begin{equation}
\mathcal{L}_{\text{total}} = 
\mathcal{L}_{\text{utt}} + \mathcal{L}_{\text{phon}} +
\mathcal{L}_{\text{CTC}} + \mathcal{L}_{\text{PCL}} +
\mathcal{L}_{\text{UCL}} + \mathcal{L}_{\text{FA}},
\end{equation}
where $\mathcal{L}_{\text{CTC}}$ (Eq.~\ref{eq:ctc}), $\mathcal{L}_{\text{PCL}}$ (Eq.~\ref{eq:pcl}), 
$\mathcal{L}_{\text{UCL}}$ (Eq.~\ref{eq:ucl}), and $\mathcal{L}_{\text{FA}}$ (Eq.~\ref{eq:fa}) 
correspond to the CTC, phoneme-level contrastive, utterance-level, 
and FA-aware losses, respectively, all of which are assigned an equal weight of 1. To maintain the focus of this study, the exploration of alternative weighting strategies is left beyond the scope of this work.
\section{Experimental Setup}
\vspace{-10pt}

\subsection{Datasets}
We use the LibriPhrase \texttt{train-clean-100} and \texttt{train-clean-360} sets with MUSAN noise~\cite{musan} for training.
Evaluation is conducted on four benchmarks: LibriPhrase Easy/Hard (L$_E$/L$_H$) from \texttt{train-other-500} (low/high phonetic confusion), Google Speech Commands V2 (G)~\cite{gsc} (35 commands under diverse conditions), Qualcomm Keyword Speech (Q)~\cite{qual} (accented/domain-specific keywords), and AMI~\cite{ami} (12h meeting recordings segmented into 2s clips for FA evaluation).
\vspace{-10pt}

\subsection{Implementation Details}
We employ Google speech embeddings~\cite{googleembedder} as the pre-trained audio encoder. 
All models are trained for 50 epochs using Adam with a fixed learning rate of $10^{-3}$, batch size $N=1000$, and mini-batch size $M=5$ for UCL (Section~\ref{sec:cl}). 
Experiments are conducted on an NVIDIA RTX 4090 GPU using TensorFlow.

\begin{figure*}[t]
    \centering
    \vspace{-15pt}  
    \includegraphics[width=1.0\textwidth, trim=0cm 0cm 0cm 0cm, clip]{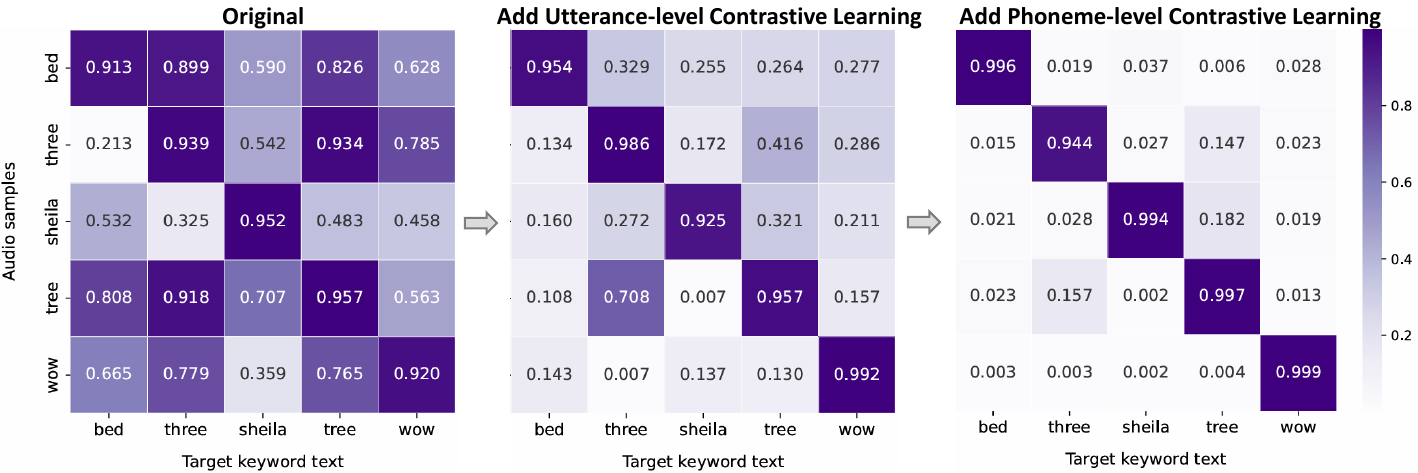}
    \vspace{-10pt}  
    \caption{Cosine similarity matrices between audio samples (y-axis) and textual keywords (x-axis) for three variants: baseline (left), UCL (middle), and UCL+PCL (right). Adding UCL reduces off-diagonal similarities and improves inter-class separation, while integrating PCL sharpens diagonal matches and suppresses residual confusions. 
    }
    \vspace{-5pt}  
    \label{fig:ucl}
\end{figure*}

\begin{figure}[t]
    \centering
    \vspace*{-8pt}  
    \includegraphics[width=0.48\textwidth, height=0.2\textheight]{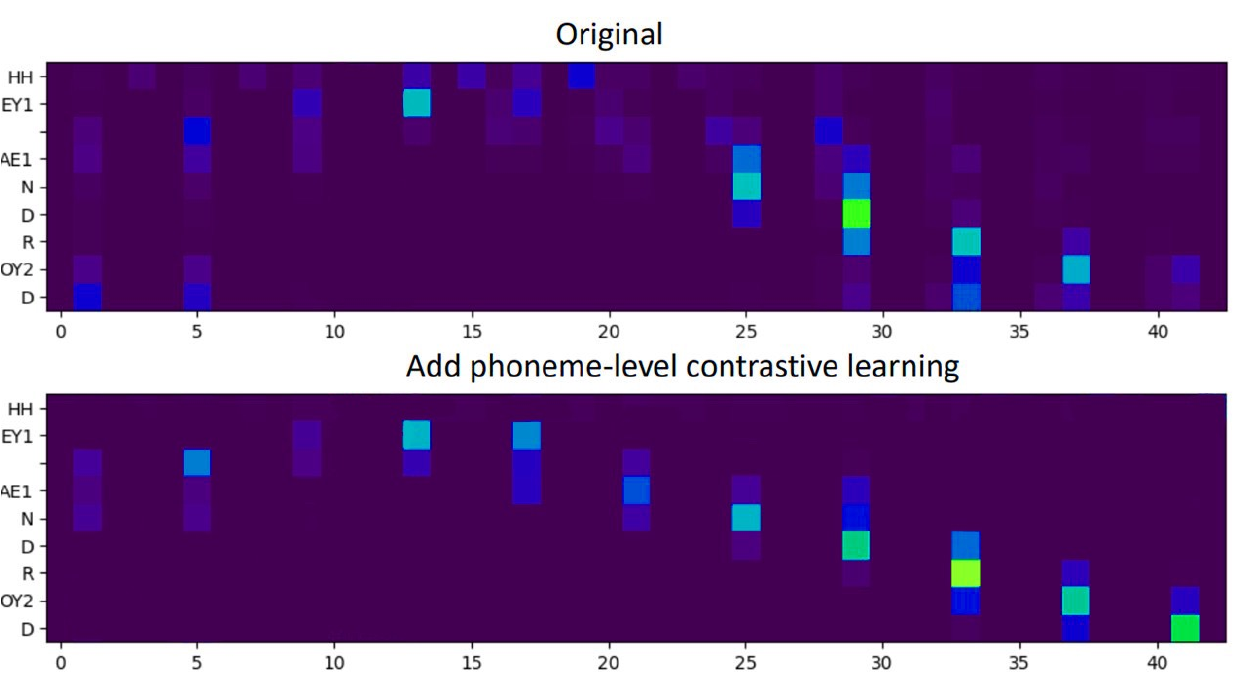}
    \caption{Phoneme-to-frame alignment heatmaps before (top) and after (bottom) applying PCL. Rows correspond to phonemes and columns to audio frames, with brighter cells indicating stronger alignment confidence. With PCL, alignments become sharper and more localized, demonstrating improved phoneme-level discrimination.
    }
    \vspace*{-8pt}  
    \label{fig:pcl}
\vspace{-10pt}

\end{figure}

\section{Experimental Results}
\subsection{Main Results}
Table~\ref{tab:performance} compares MALEFA with prior ZSKWS models and presents an ablation study. While CED~\cite{ced} achieves strong accuracy, its Conformer-based encoder~\cite{conformer} incurs much higher complexity, limiting on-device usage. Compared with PhonMatchNet~\cite{phonmatchnet}, on LP$_H$, it suffers a significant drop (AUC $=88.52$, EER $=18.82$), whereas \textbf{Ours} maintains higher robustness (AUC $=93.58$, EER $=13.91$). Ablation results show that removing FA-aware loss (\textit{w/o} FA) sharply increases EER, excluding UCL degrades robustness to phonetic ambiguities, and discarding PCL reduces fine-grained alignment on LP$_H$.  
In contrast, the full MALEFA, integrating all three components, achieving state-of-the-art performance across benchmarks with only 0.7M parameters. 
This confirms that FA-aware loss, UCL and PCL are jointly essential for reliable ZSKWS.
\vspace{-10pt}
\subsection{False Alarm Results}
Table~\ref{tab:fasle alarm} reports the FAR across test sets. 
Compared with~\cite{phonmatchnet}, which suffers from high FARs (17.9\% on AMI), our MALEFA reduces FAR to below 0.01\% on all benchmarks.  
Ablation further confirms the contribution of each component: removing FA-aware loss (\textit{w/o FA}) causes the largest degradation (14.5\% on AMI), excluding UCL increases FAR to 1.3\%, and discarding PCL still yields higher FAR (0.085\% on AMI).  
Overall, the complete MALEFA achieves the lowest FAR across all datasets, highlighting that PCL, UCL, and FA-aware learning are complementary and jointly indispensable.
\subsection{Effects of Multi-granularity Contrastive Learning on Audio-Text Matching}
To illustrate the impact of contrastive objectives, we visualize cosine similarity matrices for five representative keywords from the G dataset in Fig.~\ref{fig:ucl}. 
The baseline model (\textit{Original}) exhibits high similarity not only on correct matches but also across confusable pairs (e.g., “bed” vs. “three”), indicating risk of FA. 
Introducing UCL enforces stronger inter-class separation, effectively suppressing non-matching similarities and yielding cleaner diagonal patterns. 
Further adding PCL sharpens the alignment, driving non-matching scores close to zero while preserving near-perfect self-matches. 
These results qualitatively confirm that UCL improves global discrimination and PCL complements it with fine-grained alignment, together enhancing robustness against phonetically similar triggers.
\subsection{Effects of Phoneme-level Contrastive Learning on Frame-wise Alignment}
Figure~\ref{fig:pcl} compares phoneme-to-frame attention maps for the keyword “hey android” with and without PCL. 
Without PCL, the attention is diffused, yielding imprecise phonetic boundaries that may cause false alarm. 
With PCL, alignments become sharper and more localized, indicating that the model learns more discriminative frame-level representations and improves robustness against acoustically similar distractors.
\section{Conclusion and Future Work}
In this work, we have presented MALEFA, a lightweight ZSKWS framework that avoids reliance on large pre-trained models. 
By integrating multi-granularity contrastive learning with a novel false alarm-aware loss, MALEFA effectively captures global semantics and fine-grained pronunciations, and directly suppresses false triggers.
Experiments show that MALEFA delivers state-of-the-art performance with 99\% AUC, 1\% EER, and an ultra-low FAR of 0.007\%, making it highly suitable for resource-constrained deployments. 
In future work, we plan to explore cross-lingual extensions to improve robustness across diverse languages.
\section{ACKNOWLEDGMENTS}
This work was supported in part by Realtek Semiconductor Corporation under Grant Numbers 113KK01103 and 114KK01005. Any findings and implications in the paper do not necessarily reflect those of the sponsors.
\vfill\pagebreak

\bibliographystyle{IEEEbib}
\bibliography{main}

\end{document}